\documentclass[floatfix, noshowpacs, preprintnumbers, twocolumn, amsmath, amssymb, aps, superscriptaddress, prb]{revtex4}

\pdfoutput=1

\usepackage[english]{babel}
\usepackage{graphicx, xcolor}
\usepackage[caption=false]{subfig}
\usepackage{soul}
\usepackage{mathrsfs}
\usepackage{polski}
\usepackage{amsthm}
\usepackage{mathtools}
\usepackage{subfig}

\newcommand{\ket}[1]{\left| #1 \right>}

\newtheorem{theorem}{Theorem}[section]
\newtheorem{corollary}{Corollary}[theorem]
\newtheorem{lemma}[theorem]{Lemma}
\newtheorem{proposition}{Proposition}[section]

\theoremstyle{definition}
\newtheorem{definition}{Definition}[section]

\begin{document}

\title{Quantum Computation for Pricing Caps using the LIBOR Market Model}

\author{Hao Tang}
\altaffiliation{These authors contributed equally to this work.}
\affiliation{Center for Integrated Quantum Information Technologies (IQIT), School of Physics and Astronomy and State Key Laboratory of Advanced Optical Communication Systems and Networks, Shanghai Jiao Tong University, Shanghai 200240, China}
\email{htang2015@sjtu.edu.cn} 

\author{Wenxun Wu}
\altaffiliation{These authors contributed equally to this work.}
\affiliation{TuringQ Co., Ltd., Shanghai 200240, China}

\author{Xian-Min Jin}\email{xianmin.jin@sjtu.edu.cn}
\affiliation{Center for Integrated Quantum Information Technologies (IQIT), School of Physics and Astronomy and State Key Laboratory of Advanced Optical Communication Systems and Networks, Shanghai Jiao Tong University, Shanghai 200240, China}
\affiliation{TuringQ Co., Ltd., Shanghai 200240, China}


\maketitle

\textbf{The LIBOR Market Model (LMM) is a widely used model for pricing interest rate derivatives. While the Black-Scholes model is well-known for pricing stock derivatives such as stock options, a larger portion of derivatives are based on interest rates instead of stocks. Pricing interest rate derivatives used to be challenging, as their previous models employed either the instantaneous interest or forward rate that could not be directly observed in the market. This has been much improved since LMM was raised, as it uses directly observable interbank offered rates and is expected to be more precise. Recently, quantum computing has been used to speed up option pricing tasks, but rarely on structured interest rate derivatives. Given the size of the interest rate derivatives market and the widespread use of LMM, we employ quantum computing to price an interest rate derivative, caps, based on the LMM. As caps pricing relates to path-dependent Monte Carlo iterations for different tenors, which is common for many complex structured derivatives, we developed our hybrid classical-quantum approach that applies the quantum amplitude estimation algorithm to estimate the expectation for the last tenor. We show that our hybrid approach still shows better convergence than pure classical Monte Carlo methods, providing a useful case study for quantum computing with a greater diversity of derivatives.}

Financial products are typically classified into three categories\cite{Hull2003,Tuckman2012,Chacko2016}: equities (which typically refers to stocks), fixed income, and derivatives. The name of fixed income can be understood from its typical products, bonds, which provide investors with a fixed amount of income known as the coupon on a semi-annual basis using the risk-free market interest rate as the discount factor. Derivatives include financial instruments such as options, futures, forwards and swaps, and their underlying securities are still equity or fixed income products. Black-Scholes model\cite{Black1973, Merton1973}, one of the most well-known financial models, is raised for pricing stock options. It models the price variation of the stock with a geometric Brownian motion and then compares the instantaneous underlying stock price with the strike price. 

On the other hand, a larger portion of derivatives are those based on interest rate products instead of stocks. For instance, interest rate derivatives account for 79\% of the notional value of all derivatives in EU\cite{EUreport2021} during 2021. As an example of this type of derivatives, the interest rate cap comprises a basket of call options that protects the buyer against rises in floating interest rates above a certain nominated upper limit called cap rate. Other interest rate derivatives such as floor, collar, and swaption are also similarly defined. For decades, they have been modeled using an extension of the Black equation. Such short rate models based on instantaneous interest rates suffer from their weakness in capturing the correlation and covariance among different forward rates. In 1987, the Heath-Jarrow-Morton (HJM) model\cite{Heath1990,Heath1991,Heath1992} was introduced as a well-known instantaneous forward rate model that directly models the evolution of forward rates. However, neither of the two rates (instantaneous interest $/$ forward rates) can be directly observed in the market, which reduces the accuracy of the models. 

Nowadays, the LIBOR Market Model (LMM)\cite{Brace1997}, developed in 1997, is a more widely used model. LIBOR stands for London Interbank Offered Rate, a globally recognized benchmark interest rate at which major global banks borrow short-term loans from one another. The LIBOR rates included the rates of different maturities, also called tenors, ranging from 1 day to 12 months. It is to be noted that LIBOR is not in use starting in 2022. Now other candidates such as SONIA are used as the near risk-free interest rate benchmark in the UK (See Appendix A), and alternative interbank offered rates such as Euribor and Ameribor\cite{Tuckman2012} are also widely recognized. Nonetheless, LMM exemplifies the market model's general features, namely the utilization of readily observable market data along with the volatilities associated with traded contracts. As a result, it is far more practical than earlier pricing models of interest rate derivatives.

Essentially, LMM is described by a stochastic differential equation that has to be numerically solved by Monte Carlo simulation\cite{Huang2014,Kajsajuntti2004,Riga2011, Xiong2013, Pena2010, Brigo2006}. In recent years, quantum computing for finance applications has been a rapidly developing field\cite{Baaquie2007,Zhang2010,Meng2016, Mugel2022, Hegade2021, Stamatopoulos2021, Coyle2021, Orus2019,Rebentrost2018, Woerner2019, Martin2019, Zoufal2019, Stamatopoulos2019, Egger2019, Tang2020b, Miyamoto2022, Miyamoto2022b}, especially for replacing the comprehensively used Monte Carlo method with the quantum amplitude estimation (QAE) algorithm to achieve a quadratical speedup\cite{Brassard2002}. So far, applications of QAE on option pricing\cite{Stamatopoulos2019}, credit risk analysis\cite{Egger2019}, and collateralized debt obligation pricing\cite{Tang2020b} have been demonstrated. Considering the large market of interest rate derivatives and the wide use of LMM, it is also of great interest to apply quantum computation of the LMM to pricing interest rate derivatives, such as caps.

In this paper, we offer a quantum circuit implementation for pricing caps based on the LMM. We first introduce the principles of the LMM and caps, and then demonstrate how the stochastic differential equations associated with caps may be quantitatively solved using classical Monte Carlo methods. As caps pricing relates to path-dependent Monte Carlo iterations for different tenors, which is common for many complex structured derivatives, we develop our hybrid classical-quantum approach that applies QAE to the expectation estimation for the last tenor. We show that our hybrid approach still outperforms pure classical Monte Carlo methods in terms of convergence, providing a useful case study for quantum computing with a greater diversity of derivatives.  

\section{The LIBOR Market Model and cap pricing}

\subsection{The pricing for caps}

Given that the caps and swaptions markets are the two primary interest-rate-options markets, it is critical that a model is compatible with their market formulas. Prior to the introduction of market models, there were no interest-rate dynamics compatible with either Black's formula for caps or Black's formula for swaptions. These formulas were actually based on mimicking the Black and Scholes model for stock options under some simplified and imprecise assumptions about interest-rate distributions. The introduction of market models provided a new derivation of Black's formulas based on rigorous interest-rate dynamics.

An interest rate \textit{cap} with cap rate $K$ and tenor structure $\mathcal{T} = \{T_0, ..., T_M\}$ (and corresponding set of year fractions $\tau = \{\tau_1, ..., \tau_M\}$) is a contract which the holder of the cap receives the amount at time $T_i$
    \begin{equation}
    \label{eq:C}
        X_i = \tau_i \cdot (L(T_{i-1}, T_i) - K)^+
    \end{equation}
for each $i = 1, ..., M$, where $L(T_{i-1}, T_i)$ is simply-compounded spot interest rate as defined in Appendix~\ref{App:basics}.

Each individual $X_i$ is referred to as a \textit{caplet}. As a result, the cap is a portfolio of the individual caplets $X_1, ..., X_M$. One may have noticed that the spot interest rate $L(T_{i-1}, T_i)$ is determined already at time $T_{i-1}$. Thus, the amount $X_i$ is determined at $T_{i-1}$ but not paid until $T_i$. The market price of a cap at time $t$ is given by
\begin{equation}
\label{eq:CPP}
\begin{split}
    & Cap(t, \mathcal{T}, \tau, K) =\\
    & \sum\limits_{i = 1}^M \tau_i \mathbb{E}^Q [D(t, T_i) (L(T_{i-1}, T_i) - K)^+ | \mathcal{F}_t].   
\end{split}
\end{equation}

\begin{figure}[ht]
\centering
\includegraphics[width=0.49\textwidth]{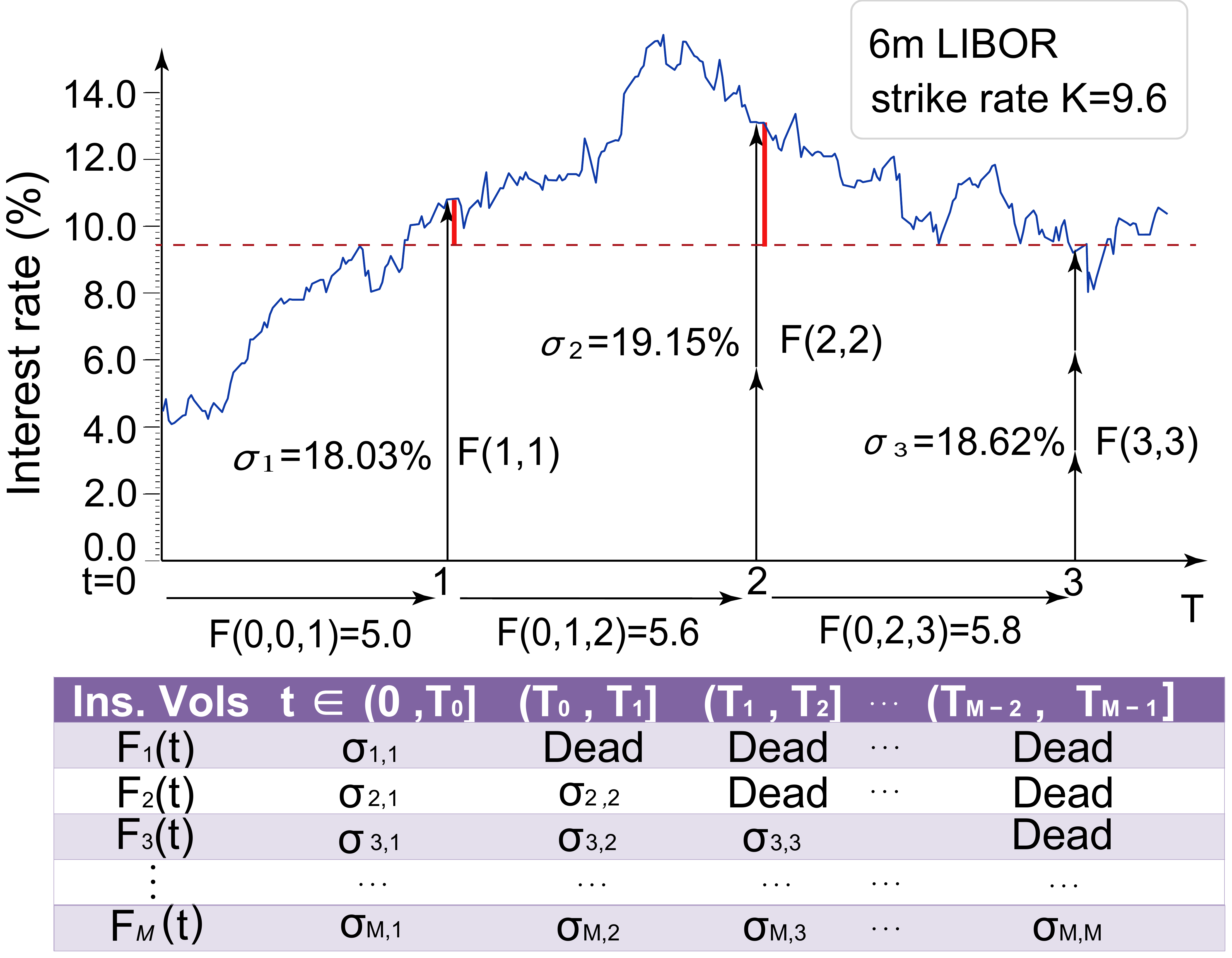}
\caption{\label{fig:G1}\textbf{Illustration of the Payoff of a Sample Cap and a Simulated LIBOR Path.} The blue line starting at an interest rate value of 4.69\% is a simulated 6-month LIBOR path. The dotted red line indicates the strike rate of an interest rate cap which pays off at time t=\{1, 2, 3\}. If 6-month LIBOR exceeds the strike rate, the buyer receives a certain amount of the payoff, as shown in vertical red lines. Vertical arrows demonstrate how 6-month LIBORs are simulated using forward rates and implied volatility. The table below shows the structure of the implied volatility, with the assumption that implied volatility is piecewise constant, i.e., constant in rows.}
\end{figure}

The market practice is to use the so called Black-76 formula for the time $t$ pricing of caplets. In this case, $\mathbb{E}^Q$ means that the expectation is taken under risk-neutral measure $Q$. Fig.~\ref{fig:G1} shows an example of how a cap's payoff is calculated. A 6-month USD LIBOR cap with a cap rate of $K = 9.6$ and a tenor structure of $\mathcal{T} = \{1,\; 2,\; 3\}$ is depicted in the figure. If the 6-month USD LIBOR is greater than the strike rate of $K = 9.6$ on any of the dates in the set $\mathcal{T}$, cap holder will be paid with amount $X_i = R_i - K$ times the notional value of the cap, where $R_i$ is the 6-month USD LIBOR on that date. Dates of potential payoff (and the amount of the payoff) are colored red in Fig.~\ref{fig:G1}.

The Black-76 formula for the caplet defined in Eq.~\ref{eq:C} is given by
\begin{equation}
    Capl^{Black}(t, T_{i-1}, T_{i}, K, v_i) = \tau_i p(t, T_i) Bl(K, F_i(t), v_i)
\end{equation}
where
\begin{equation}
\begin{split}
    Bl(K, F_i(t), v_i) & = F_i(t) \Phi(d_1(K, F_i(t), v_i))\\ 
    & - K \Phi(d_2(K, F_i(t), v_i)),\\
    d_1(K, F, u) & = \frac{ln(F/K) + u^2/2}{u},\\
    d_2(K, F, u) & = \frac{ln(F/K) - u^2/2}{u},\\
    v_i & = v\sqrt{T_{i-1} - t},\\
\end{split}
\end{equation}
$v$ is the volatility retrieved from market quotes. $\Phi$ is the standard Gaussian cumulative distribution function.

The price of the cap at time $t$ is
\begin{equation}
    Cap^{Black}(t, \mathcal{T}, \tau, K, v) = \sum\limits_{i = 1}^M Capl^{Black}(t, T_{i-1}, T_{i}, K, v_i).
\end{equation}

The problem is that, when we discount the value of the caps, we assume the discount factor $p(t, T_i)$ to be deterministic and identify it with the corresponding bond price. But when we price caplets using the traditional Black-Scholes method, we assume that the forward LIBOR rate moves in a Brownian way.

This inconsistency can be solved using the change of numeraire technique. The LIBOR market model approach is to first assume that the LIBOR forward rates have the dynamics 
\begin{equation}
    \label{eq:1c}
    d F_i(t) = \sigma_i (t) F_i (t) d Z_i^i (t),
\end{equation}
where $Z_i^i$ is the $i$-th component of the $Q^i$-Wiener as described in Appendix~\ref{App:basics}. Then we have a discrete tenor LIBOR market model with volatilities $\sigma_i, ..., \sigma_M$. We demonstrate how the dynamics of each $F_i$ are correlated In theorem~\ref{thm:FMDL}. 

Assuming the LIBOR model exists, we can change Eq.~\ref{eq:CPP} from the risk-neutral measure $Q$ with numeraire $B$ to the $T_i$-forward measure in each $i$-th summand\cite{Girsanov1960}. Then we have summation
\begin{equation}
    \sum\limits_{i = 1}^M \tau_i p(t, T_i) \mathbb{E}^i [(L(T_{i-1}, T_i) - K)^+ | \mathcal{F}_t]
\end{equation}
instead. Appendix~\ref{App:basics} contains the primary findings that were utilized to demonstrate the equivalency. Appendix~\ref{App:Ito}, Appendix~\ref{App:p1.2}, and Appendix~\ref{App:pER} provide further evidence and information.

\begin{proposition}
    In the LIBOR market model, each individual caplet price in Eq.~\ref{eq:CPP} is given by
    \begin{equation}
        Capl^{LMM} (t, T_{i-1}, T_i, K) = Capl^{Black} (t, T_{i-1}, T_i, K, v_i),
    \end{equation}
    where
    \begin{equation}
        v_i^2 = \int_t^{T_{i-1}} \sigma_i(t)^2 dt.
    \end{equation}
\end{proposition}

Next we will show how $\sigma_i(t)$ is calibrated under LIBOR market model.

\subsection{Calibration of the LMM Model to Caps}

First, one must note that the joint dynamics of forward rates are not involved in the pricing of caps, as there are no expectations involving two or more forward rates at the same time. As a result, the correlation between different rates does not appear in our pricing formula. However, if we wish to price more complex interest rate derivatives such as Bermudan swaptions, whose evaluation will be based on expected values of quantities involving several rates at the same time, we must take the correlation between each Wiener process into account. Pricing these derivatives is left for our future work.

$v_{T_{i - 1}-caplet}^2$ is defined as the average instantaneous variance over time, i.e. $v_i^2$ normalized with respect to time. So we have
\begin{equation}
    v_{T_{i - 1}-caplet}^2 = \frac{1}{T_{i - 1}} \int_t^{T_{i-1}} \sigma_i(t)^2 dt.
\end{equation}

$v_{T_{i - 1}-caplet}$ is thus the square root of the average percentage variance of the forward rate $F_i(t)$ for $t \in [0, T_{i - 1})$ and is called $T_{i-1}$-caplet volatility.

In order to calibrate the model, we have to make an assumption. We assume that the instantaneous volatility of the forward rate $F_k(t)$ is piecewise-constant, as is commonly assumed in practice\cite{Huang2014, Brigo2006}. Under this assumption, it is possible to organize instantaneous volatilities in the table in Fig.~\ref{fig:G1}.

We consider the forward interest rate and implied volatility provided in the book by Brigo and Mercurio\cite{Brigo2006} as our benchmark dataset. Fig.~\ref{fig:G1} shows how the LIBOR rate is simulated under the LIBOR market model. At $t = 0$, the observed LIBOR rate is $4.69\%$. Like is assumed in Brigo and Mercurio, 2006, for each $F_i(t), i = 1, 2, \dots$, the implied volatility is piecewise constant, i.e., $\sigma_{i, 1} = \sigma_{i, 2} = \dots = \sigma_{i, j}$ for any $j \in \{1, \dots, i\}$.  LIBOR at each checking time point $t = 1, 2, \dots$ can be properly simulated using observed forward interest rates $\{F(0, 0, 1), F(0, 1, 2),\dots\}$ and implied volatilities.

\subsection{Monte Carlo Simulation}

In this section, we use the parameters in the benchmark dataset to construct a Monte Carlo simulation to price a 3-year cap. As we have already shown in earlier subsections, under $T_i$-forward measure, each $L(T_{i-1}, T_i) = F_i(T_{i-1})$ is just a GBM. We need to discretize time from $t$ to $T_M$. We choose $t = T_0$ and each $T_i - T_{i-1}$ to be a year. From Eq.~\ref{eq:GBMUIL}, if we evolve the forward rate vector directly from one tenor date to the next, we have
\begin{equation}
    \hat{F_i}(T+\Delta) = \hat{F_i}(T)\cdot e^{\sqrt{\Delta} \sigma_i Z_i^i - \frac{1}{2}\sigma_i^2 \Delta},
\end{equation}
where $\Delta = 252/252 = 1$, 252 is the average number of working days in a year, and $Z^i_i$ has a standard Gaussian distribution.

\section{The quantum circuit construction}

\begin{figure*}[ht]
    \centering
    \includegraphics[width=0.95\textwidth]{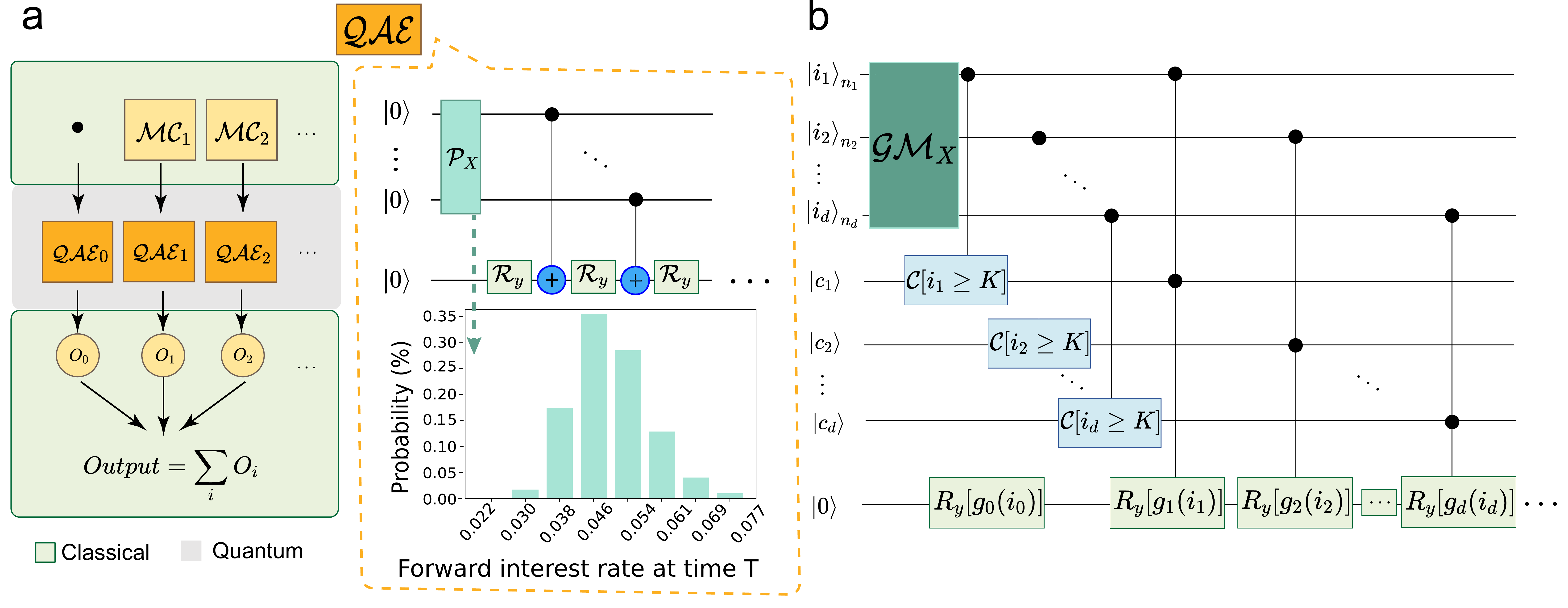}
    \caption{\label{fig:hcqc}\textbf{Quantum-Classical Hybrid Method and Pure Quantum Circuit to Perform Caps Pricing.} \textbf{(a)} The hybrid Quantum-classical framework, where $\mathcal{MC}_i$ represents the $i$-th classical Monte-Carlo simulation, $\mathcal{O}_i$ reperesents the simulated final payoff of the $i$-th caplet, and $\mathcal{QAE}$ represents the quantum circuit for payoff measurement. The log-normal distribution is loaded on a 3-qubit quantum register using the operator $\mathcal{P}_X$ inside $\mathcal{QAE}$. The payoff is captured by $\mathcal{R}_y$ gates, whose expected value is measured by the QAE algorithm. \textbf{(b)} The pure quantum circuit framework, in which the operator $\mathcal{GM}_X$ uses a number of qubits to load the multivariate distribution of multiple caplets ($3 + 6 + 9 = 18$ qubits in our example). Each quantum register $i_d$ represents for one caplet. $\mathcal{C}[i_d \geq K]$ is the comparator circuits which set $\ket{c_i}$ to $\ket{1}$ if the value of $i$-th caplet is larger than the strike rate.}
\end{figure*}

We use quantum computing as an alternative to Monte Carlo simulations for cap pricing. Many structural interest rate derivatives, including caps, share a common problem in that they require path-dependent Monte Carlo iterations for various tenors. In this section, we present three distinct methods for evaluating the caps under the LIBOR Market Model. We set the Black-76 formula calculated outcome as the benchmark.
\begin{itemize}
    \item Classical Monte Carlo method: we use Monte Carlo simulations to price every caplets and compute the price of caps.
    \item Quantum-classic hybrid method: we use Monte Carlo simulations to simulate forward rates until the second-last alive one. Using a simulated forward path, we compute the expected payoff using QAE. 
    \item Pure Quantum method: we make use of QAE to compute the expected payoff of the entire cap. We need separate qubits to load the probability distribution for each caplet.
\end{itemize}

The framework for the Quantum-classic hybrid method and Pure Quantum method are shown in Fig.~\ref{fig:hcqc} (a) and Fig.~\ref{fig:hcqc} (b) respectively. $\mathcal{MC}_i$ module in Fig.~\ref{fig:hcqc} (a) is classical simulation method that simulates the forward rate value until second-last alive one. The simulated result is then passed to the $\mathcal{QAE}_i$ module, which performs a quantum amplitude estimation to determine the expected payoff. In the end, a classical circuit sums up the result of each caplet. The unitary $\mathcal{P}_x$ represents the set of gates that load the random distribution. $\mathcal{P}_x$ and $\mathcal{QAE}_i$ will be discussed in the sections that follow.

\subsection{Load the random distribution}
The random distribution for the possible asset prices in the future can be loaded on a quantum register\cite{Egger2019}. In the scenario of hybrid method for caps pricing, we load the uncertainty of the forward rate in year 3 into the quantum state. Each basis state represents a possible value, and its amplitude represents the corresponding probability. The distribution loading module creates the following entangled states:
\begin{equation}
    |\Psi\rangle_n = \sum_{i = 0}^{2^n - 1} \sqrt{p_i} |S_i\rangle_n.
\end{equation}

\subsection{Calculate expected payoff using QAE}
QAE has been demonstrated as a good alternative to Monte Carlo simulation for financing products.\cite{Rebentrost2018, Stamatopoulos2019, Woerner2019, Egger2019} QAE can achieve quadratic speedup, but involvement of inverse Quantum Fourier Transform (QFT) requires exponentially increasing circuit depths. So we implement an iterative QAE (IQAE) \cite{Qiskit2019} for pricing caplets.

For the Quantum-classic hybrid method: we use Monte Carlo simulations to simulate forward rates until the second-last alive one. Using a simulated forward path, we compute the expected payoff using QAE. Each probability distribution is loaded using $n = 3$ qubits.

\begin{figure*}[ht]%
    \centering
    \includegraphics[width=0.95\textwidth]{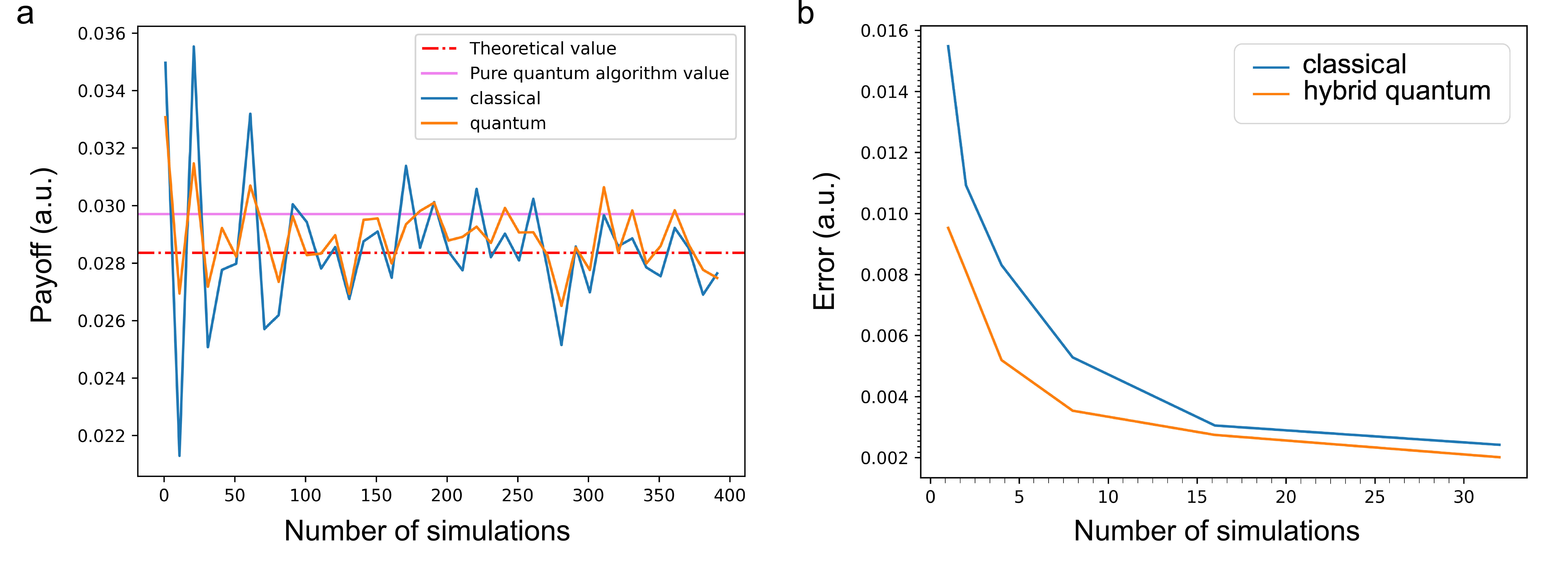}
\caption{\label{fig:E}\textbf{Caps rate results and the estimation error.} (a) shows the pricing results using different methods. The theoretical value is calculated using the Black-76 formula under the LIBOR market model settings. The number of qubits used to construct the probability distribution in the pure-quantum method is 18. (b) demonstrates the absolute estimation error of the classical Monte Carlo method and the quantum-classic hybrid method, averaged over 50 trails. The x-axis indicates the number of simulations of the Monte Carlo method. The quantum-classic hybrid method uses 3 qubits to load the distribution in this case. }%
\end{figure*}

For the pure quantum method, we use QAE to compute the expected payoff of each caplet and sum them up. To achieve the same precision for each caplet's pricing, we use $n = 3i$ qubits to load the probability distribution for each caplet, depending on the caplet's maturity date, $T_i$. If we assume there are 8 different states at time $t = 1$, the state number must be $64 (=8 \times 8) $to achieve the same accuracy at time $t = 2$. The operator $\mathcal{GM}_X$ loads the multivariate distribution of the three caplets using a total of 18 qubits, as shown in Fig. 2(b).

\section{The result anaysis}

Fig.~\ref{fig:E} (a) shows the evaluation result under different numbers of samples for three different estimation methods. When there are fewer simulations, the pure quantum technique is more accurate than both the quantum-classical hybrid method and the classical Monte Carlo method. The quantum-classic hybrid approach better converges to the theoretical value as the number of simulations increases. It is much more difficult to improve the accuracy of the pure quantum technique due to the enormous circuit depth and number of qubits used.

The number of qubits needed to construct the pure quantum circuit with different $n$ and $T$ values is shown in Table~\ref{tab:qc}. When $n \geq 4$ and $T = 3$, we require a large number of qubits ($\geq 48$) to build the circuit. The pure quantum circuit approach for QAE in derivative pricing requires a certain amount of computational resource\cite{Chakrabarti2021} either using a quantum computing simulation framework like Qiskit or using quantum computers. However, the result will not improve much per se. 

\begin{table}[ht]
\centering
\caption{Number of Qubits Needed to Construct the Pure Quantum Circuit for Distribution Loading Qubits $n$ and Duration $T$}
\begin{tabular}{ l p{1cm} c } \hline
Gate & & logical qubits\\
\hline
Distribution loading & & $\frac{(n+nT)T}{2}$\\
Comparator & & $\frac{(n+nT)T}{2}$ + 1\\
Y-rotations & & 1\\
\hline
Quantum Amplitude Estimator & & m*\\
\hline
Total & & $\approx nT^2 + nT$ \\
\multicolumn{3}{c}{\footnotesize *: m depends on the number of sampling qubits in QAE part.}
\end{tabular}
\label{tab:qc}
\end{table}

Fig.~\ref{fig:E} (b) shows the average absolute deviation of the classical Monte Carlo method and the quantum-classic hybrid method from theoretical value over 50 trials. The quantum-classic hybrid method has consistently lower estimation errors and thus can speed up the estimation process.

Under the assumption that sample standard deviation of Monte Carlo simulation option prices is unchanged\cite{Jabbour2011}, the Central Limit Theorem can be used to describe the size of Monte Carlo simulation errors of any caplet of time length $T$ years, when the simulation number is large:
\begin{equation}
\label{eq:MCC}
    \epsilon_M[Capl_i] \approx \sigma \nu \frac{1}{\sqrt{M}} 
\end{equation}
where $\sigma_i$ denotes the sample standard deviation of Monte Carlo simulation option prices, $M$ denotes the number of Monte Carlo simulations, and $\nu$ denotes a standard normal random variable. While the estimation error of QAE with $M$ quantum samples is given by\cite{Stamatopoulos2019}:
\begin{equation}
\label{eq:QAEC}
    \epsilon_M[Capl_i] \leq \frac{\pi}{M} + \frac{\pi^2}{M^2}
\end{equation}

For any 1-year time step, we do $N$ Monte Carlo simulations. To estimate the caplet price for a caplet with a time length of $T$ years, we run $N^T$ simulations. One can see the theoretical explanation of the reduction of estimation error in the quantum-classic hybrid method by comparing Eq.~\ref{eq:MCC} and Eq.~\ref{eq:QAEC}.

\begin{figure*}[ht]
\centering
\includegraphics[width=0.9\textwidth]{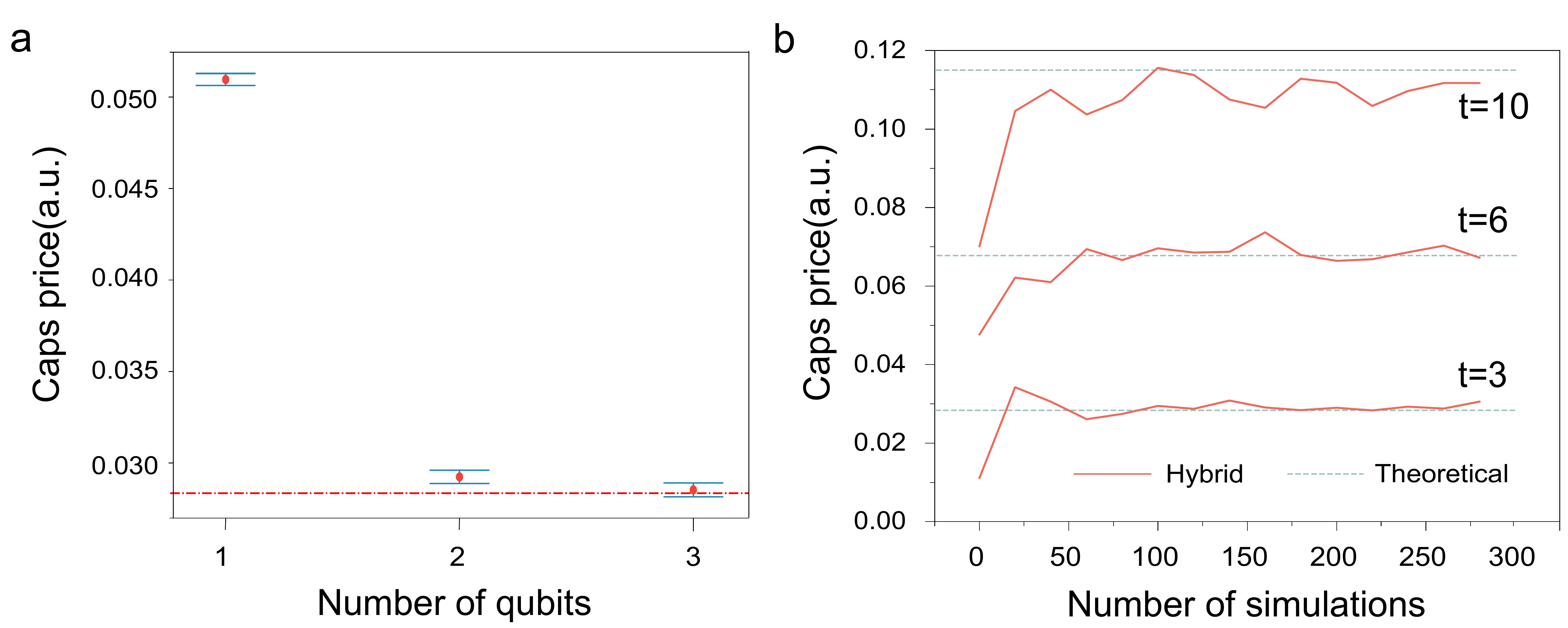}
\caption{\textbf{Simulation Results for Various Numbers of Qubits and Time Lengths.} (a) shows the sample mean and confidence interval under different numbers of sample qubits. The x-axis indicates the number of quibts used to construct the log-normal distribution. The y-axis is the price of the option. Red dots show the sample mean of 100 experiments with different qubit number $n$s. Error bars represent standard errors of the mean. The red dotted line is the theoretical value. (b) shows the outcome of the quatum-classic hybrid method and the classical theoretical method for caps under different $T$ values. Data is from Brigo and Mercurio, 2006\cite{Brigo2006}. The x-axis represents the number of simulations of the Monte-Carlo method, whereas the y-axis represents the value of the pricing outcome. The dotted line is the theoretical value of the caps and is calculated using the Black-76 formula under LIBOR maraket model setups.}
\label{fig:CI}
\end{figure*}

In the above example, we use $3T$ qubits to load the probability distribution of caplets with maturity $T$. Changes in the number of qubits used to load the target probability distribution can affect the final outcomes. In Fig.~\ref{fig:CI} (a) we compare the impact of different numbers of qubits on the final pricing result. Number $n = 1, 2, 3$ means the the probability distribution of caplets with maturity $T$ being loaded using $nT$ qubits. Mean and $95\%$ confidence interval of each different $n$ are calculated from 50 replicated experiments of pure-quantum circuits. The standard deviations of different $i$ values are in fact low and do not change much (in our experiment, standard deviation $= 1.41 \times 10^{-3}$, $1.55\times 10^{-3}$, and $1.65\times 10^{-3}$, respectively). The red line shows the theoretical value. The theoretical value is not covered by the $95\%$ confidence interval when $n = 2$, but the sample mean ($2.85\times 10^{-2}$) is close to the theoretical value ($2.84\times 10^{-2}$) within the $95\%$ confidence interval when $n = 3$.

 The results of the quantum-classic hybrid method with $n = 3$ are compared to theoretical values for various cap lengths of $T$ in Fig.~\ref{fig:CI} (b). The convergence of the hybrid method to theoretical values is clearly visible in this figure for both $T = 3$ and $T = 6$, except for $T = 10$ where there is a small divergence. One possibility is that as $T$ increases, the probability distribution approximated by $n = 3$ qubits becomes insufficiently precise. This can be further improved with gradually increasing $n$. We put a detailed illustration in Appendix~\ref{App:d}. 


\section{Discussion and Conclusion}
The LIBOR market model has become one of the most important models for pricing interest rate derivatives. It helps to solve the inconsistency between the determined discount factor and the stochastic forward LIBOR rate in the classical Black-Scholes option pricing setup. The lognormal LIBOR market model prices caps with the Black 76 formula, which is the standard formula used in the cap market. We demonstrated that by including quantum circuits into the classical Monte-Carlo method, one can accelerate convergence to the correct cap price under the LIBOR market model. 

Now a growing number of quantum techniques are being employed for financial applications. Very recent progresses include the use of quantum circuit Born machines\cite{Zhu2021, Zhu2022, Kiss2022} to load various distributions and even joint distributions such as copulas, which are widely investigated in finance. These techniques might be useful for loading multivariate distributions for these interest rate derivatives like caps. 

Furthermore, after our demonstration of the first QAE application on interest rate derivatives in this work, we suggest the LIBOR market model can be further developed to price swaptions with Black's swaption formula. These caps and swaptions are the two main kinds of interest rate derivatives. We provide a useful case study with good generalizability. It may inspire quantum computing for a larger diversity of interest rate derivatives, and can be generalized to broader cases in which a quantum-classic hybrid solution is required.

\section*{Acknowledgement.}
H.T. thanks Prof. Stephen Schaefer's previous supervision on fixed income and interest rate derivative research at London Business School. This research was supported by National Key R\&D Program of China (2019YFA0308703, 2017YFA0303700), National Natural Science Foundation of China (61734005, 11761141014, 11690033, 11904229), Science and Technology Commission of Shanghai Municipality (STCSM) (21ZR1432800,17JC1400403), and Shanghai Municipal Education Commission (SMEC) (2017-01-07-00-02- E00049). H.T. and X.-M.J. acknowledge additional support from a Shanghai talent program.

\section*{Author Contributions} 
H.T. and X.-M.J. conceived and supervised the project. H.T. and W.W. designed the scheme. W.W. wrote the quantum computing code and did Monte Carlo simulation. H.T., W.W. and X.M.J. analyzed the data and presented the figures. H.T., W.W. wrote the paper, including the appendix, with input from all the other authors.

\section*{Competing Interests.}
The authors declare no competing interests.

\section*{Data Availability} 
The data that support the plots within this paper and other findings of this study are available from the corresponding author upon reasonable request.

\begin{appendix}
\clearpage
\newpage
\onecolumngrid

\setcounter{table}{0}
\setcounter{equation}{0}
\setcounter{figure}{0}
\setcounter{section}{0}

\renewcommand{\thetable}{\arabic{table}}
\renewcommand{\theequation}{{A}\arabic{equation}}
\renewcommand{\thefigure}{{A}\arabic{figure}}

\bigskip
\section{Some notes about the alternatives to LIBOR}
The LIBOR rate is the cost for leading banks to borrow from each other, and has been widely used as the benchmark risk-free rate since the 1970s. However, due to a scandal in 2012 that suggested the Barclay Bank knowingly made misleading statements relating to benchmark-setting, finance industry practitioners are now encouraged to use alternative risk-free benchmark rates, and banks are not required to submit LIBOR quotes after 2021, as announced by the Financial Conduct Authority. This means that LIBOR may still exists then, but its viability as a benchmark rate will possibly be reduced.

The Sterling Overnight Interbank Average Rate (SONIA) is another commonly used benchmark rate. SONIA is the weighted average rate of unsecured overnight transactions in the British sterling market brokered by Wholesale Markets Brokers' Association members. SONIA was established in 1997 and is calculated at every business day in London. SONIA has especially well-known application in the sterling Overnight Indexed Swap market.
 
In 2018, the Bank of England took over the duties of calculation and publication of SONIA rates from the Wholesale Markets Brokers' Association, and set stringent requirement for the regulations of SONIA rats. SONIA is now generally regarded as a preferred candidate to replace LIBOR as a near risk-free interest rate benchmark.

\section{Basics of LIBOR Market Model and Cap Pricing}
\label{App:basics}

\renewcommand{\theequation}{{B}\arabic{equation}}
\renewcommand{\thefigure}{{B}\arabic{figure}}

\subsection{The LIBOR Market Model}
We first make some definitions

\begin{definition}[Zero-coupon bond]
    A \textit{zero-coupon bond} with maturity date $T$, briefly called $T$-bond, is a contract which guarantees the holder to be paid $1$ unit of currency at time $T$, with no intermediate payments. The contract value at time $t < T$ is denoted by $p(t, T)$.
\end{definition}

\begin{definition}[Bank account]
    The \textit{bank account} at time $t \geq 0$ is the value (the $t$-value) of a bank account with a unitary investment at initial time 0. It is denoted by $B(t)$ and its dynamics is given by:\begin{equation}
        d B(t) = r(t) B(t) dt, \; B(0) = 1,
    \end{equation}
    or\begin{equation}
        B(t) = e^{\int_0^t r(s) ds}.
    \end{equation}
\end{definition}

The bank account is the only asset in the market which is not modified when moving to a risk-adjusted probability measure. It provides us with a model of the time value of money and allow us to build a discount factor of the value.

\begin{definition}[Stochastic discount factor]
    The \textit{stochastic discount factor} between time $t$ and $T \geq t$ is the amount of money at time $t$ equivalent (according to the dynamic of $B(t)$) at 1 unit of currency at time $T$. It is denoted by $D(t, T)$ and is defined by\begin{equation}
        D(t, T) = \frac{B(t)}{B(T)} = e^{-\int_t^T r(s) ds}.
    \end{equation}
\end{definition}

\begin{definition}[Absolutely continuous and equivalent measure]
    Consider a measurable space $(X, \mathcal{F})$ on which there are defined two separate measures $P$ and $Q$. If for all $A \in \mathcal{F}$, it holds that\begin{equation*}
        P(A) = 0 \Rightarrow Q(A) = 0
    \end{equation*}
    then $Q$ is said to be \textit{absolutely continuous} with respect to $P$ on $\mathcal{F}$ and we write this as $Q << P$.
    If we have both $P << Q$ and $Q << P$, then $P$ and $Q$ are said to be \textit{equivalent} and we write $P \sim Q$.
\end{definition}

\begin{definition}[Equivalent martingale measure (EMM)]
    An EMM $Q$ with numeraire $B$ is a probability measure on $(\Sigma, \mathcal{F}, P)$ such that:
    \begin{enumerate}
        \item Q is equivalent to P;
        \item the process of the discounted prices $\tilde{S} = (\tilde{S}_t)_{t \in [0, T]}$ defined by\begin{equation*}
            \tilde{S}_t = \frac{S_t}{B_t}
        \end{equation*}
        is a strict Q-martingale.
    \end{enumerate}
\end{definition}

\begin{definition}[Simply-compounded spot interest rate]
    The \textit{simply-compounded spot interest rate} at time $t$ for the maturity $T$ is the constant rate at which an investment of $p(t, T)$ unit of currency at time $t$ accrues proportionally to the investment time to produce 1 unit of currency at time $T$. It is denoted by $L(t, T)$ and is defined by:
    \begin{equation}
        L(t, T) \coloneqq \frac{1 - p(t, T)}{\tau(t, T) p(t, T)}
    \end{equation}
    where $\tau(t, T)$, the measure of time to maturity, is referred to as the year fraction between the dates $t$ and $T$ and it is usually expressed in years.
\end{definition}

\begin{definition}[Correlated Brownian motion]
    A $d$-dimensional \textit{correlated Brownian motion} $W = (W^1, ..., W^d)$ on $(\Omega, \mathcal{F}, P)$ with the filtration $(\mathcal{F}^W_t)_{t \in [0, T]}$ is defined by
    \begin{equation}
        W_t = A \cdot \bar{W}_t
    \end{equation}
    where $\bar{W}$ is a standard $d$-dimensional Brownian motion and $A = (A^{ij})_{i,j=1, ...,d}$ is a non-singular $d \times d$ constant matrix. The instantaneous correlation matrix $\rho$ is defined as
    \begin{equation}
        \rho = AA^*
    \end{equation}
    where $A^*$ is the conjugate transpose of A. And we assume that $\rho^{ii} = 1 \; a.s.$
\end{definition}

We setting up the model:
\begin{enumerate}
    \item $t = 0$ is the current time;
    \item the set $\{T_0, ..., T_M\}$ of expiry-maturity dates (expressed in years) is the tenor structure, with the corresponding fractions $\{\tau_0, ..., \tau_M\}$, i.e., $\tau_i$ is the one associated with the expiry-maturity pair $(T_{i-1}, T_i)$, for all $i > 0$, and $\tau_0$ from now to $T_0$;
    \item set $T_{-1} \coloneqq 0$;
    \item the simply-compounded forward interest rate resetting at its expiry date $T_{i - 1}$ and with maturity $T_i$ is denoted by $F_i (t) \coloneqq F(t; T_{i-1}, T_i)$ and is alive up to time $T_{i-1}$, where it coincides with the spot LIBOR rate $F_i(T_{i-1}) = L(T_{i-1}, T_i)$, for $i = 1, ..., M$;
    \item $Q_i$ is the EMM associated with the numeraire $p(\cdot, T_i)$, i.e., the $T_i$-forward measure;
    \item $Z^i$ is the M-dimensional correlated Brownian motion under $Q^i$, with instantaneous correlation matrix $\rho$.
\end{enumerate}

The focal point of the LIBOR models is the following simple result.

\begin{lemma}
For every $i = 1, ..., N$, the LIBOR process $F_i$ is a martingale under the corresponding forward measure $Q^i$, on the interval $[0, T_{i - 1}]$.
\end{lemma}

\begin{proof}
From the definition of forward rates we have 
\begin{equation}
\label{eq:FR}
\begin{split}
    F_i(t) p(t, T_i) & = \frac{p(t, T_{i - 1}) - p(t, T_i)}{\tau_i} \\
    \tau_i F_i(t) & = \frac{p(t, T_{i - 1})}{p(t, T_i)} - 1
\end{split}
\end{equation}
$1$ is obviously a martingale under any measure. The process $p(t, T_{i - 1})/p(t, T_i)$ is the price of the $T_{i-1}$ bond normalized by the numeraire $p(t, T_i)$. Since $p(t, T_i)$ is the numeraire for the martingale measure $Q^i$, the process $p(t, T_{i - 1})/p(t, T_i)$ is thus trivially a martingale on the interval $[0, T_{i - 1}]$. Thus $\tau_i F_i(t) = p(t, T_{i - 1})/p(t, T_i) - 1$ is a martingale and hence $F_i$ is also a martingale.
\end{proof}

\begin{definition}
    If the LIBOR forward rates have the dynamics\begin{equation}
    \label{eq:1}
        d F_i(t) = \sigma_i (t) F_i (t) d Z_i^i (t)
    \end{equation}
    where $Z_i^i$ is the $i$-th component of the $Q^i$-Wiener as described above, then we say that we have a discrete tenor LIBOR market model with volatilities $\sigma_i, ..., \sigma_M$.
\end{definition}

Notice that if $\sigma_i$ is bounded, the Eq.~\ref{eq:1} has a unique strong solution since it is just a Geometric Brownian Motion (GBM). Using Ito's formula in Appendix~\ref{App:Ito} on $d \ln F_i(t)$ one can easily derive
\begin{equation}
\label{eq:GBMUIL}
\begin{split}
    & d \ln F_i(t) = \sigma_i (t) d Z_i^i (t) - \frac{\sigma_i (t)^2}{2} dt\\
    & \ln F_i(T) = \ln F_i(t) + \int_t^T \sigma_i (s) dZ_i^i (s) - \int_t^T \frac{\sigma_i (s)^2}{2} ds\\
    & F_i(T) = F_i(t) e^{\int_t^T \sigma_i(s) d Z_i^i (s) - \frac{1}{2} \int_t^T \frac{\sigma_i (s)^2}{2} ds}
\end{split}
\end{equation}
for all $0 \leq t \leq T \leq T_{i - 1}$.

Next theorem shows the dynamics of each $F_k$ under the forward measure $Q_i$ different from $Q_k$.
\begin{theorem}[Forward measure dynamics in the LMM]
\label{thm:FMDL}
Under the assumptions of the LIBOR market model, the dynamics of each $F_k$, for $k = 1, ..., M$, under the forward measure $Q^i$ with $i \in \{1, ..., M\}$, is:
    \begin{equation}
        dF_k (t)=
        \begin{cases}
            F_k (t)\sigma_k(t)dZ^i_k(t) & \\
            - F_k (t)\sum\limits_{j=k+1}^i \frac{\tau_j \rho_{k, j} \sigma_k(t) \sigma_j(t) F_j(t)}{1+\tau_j F_j(t)} dt & k < i\\
            F_k (t)\sigma_k(t)dZ^i_k(t) &  k = i\\
            F_k (t)\sigma_k(t)dZ^i_k(t) & \\
            + F_k (t)\sum\limits_{j=i+1}^k \frac{\tau_j \rho_{k, j} \sigma_k(t) \sigma_j(t) F_j(t)}{1+\tau_j F_j(t)} dt & k > i\\
        \end{cases}
    \end{equation}
for $t \leq \min\{T_{k-1}, T_i\}$
\end{theorem}
\begin{proof}
    See Appendix~\ref{App:p1.2}
\end{proof}

We can easily turn Theorem~\ref{thm:FMDL} around and have the following existence result.
\begin{theorem}
\label{thm:ER}
    Consider a given volatility structure $\sigma_1, ..., \sigma_N$, where each $\sigma_i$ is assumed to be bounded, a probability measure $Q^M$ and a correlated $Q^M$-Wiener process $W^M$. Define the processes $F_1, ..., F_N$ by
    \begin{equation}
    \label{eq:PF}
    \begin{split}
        d F_i(t) = -\sigma_i(t) F_i(t) \sum\limits_{j=i+1}^M \frac{\rho_{i, j}\tau_j\sigma_j(t)F_j(t)}{1 + \tau_j F_j(t)} dt \\
        + \sigma_i(t) F_i(t) d Z_i^M(t),        
    \end{split}
    \end{equation}
    then the $Q^i$-dynamics of $F_i$ is given by Eq.~\ref{eq:1}. Thus there exists a LIBOR model with the given volatility structure.
\end{theorem}

\begin{proof}
    See Appendix~\ref{App:pER}
\end{proof}

\section{Ito's Formula}
\label{App:Ito}

\renewcommand{\theequation}{{C}\arabic{equation}}
\renewcommand{\thefigure}{{C}\arabic{figure}}

Let $X$ be a stochastic process and suppose there exists a real number $a$ and two adapted process $\mu$ and $\sigma$ such that the following relation holds for all $t \geq 0$:
\begin{equation}
\label{eq:SP}
    X_t = a + \int_0^t \mu_s ds + \int_0^t \sigma_s d W_s
\end{equation}
where $W_s$ is a Wiener process. We can write Eq.~\ref{eq:SP} in the following form:
\begin{equation}
\begin{split}
    & d X_t = \mu_t dt + \sigma_t dW_t\\
    & X(0) = a
\end{split}
\end{equation}

Assume furthermore that we have a $C^{1, 2}$-function
\begin{equation*}
    f : R_+ \times R \rightarrow R
\end{equation*}
of time $t$ and stochastic process $X_t$.

We now ask what the local dynamics of the function $f(t, X_t)$ look like. We know the answer for the nonstochastic case: Given $\sigma_t \equiv 0$,
\begin{equation}
\begin{split}
    & d X_t = \mu_t dt \\
    & d f(t, X_t) = \frac{\partial f}{\partial t} dt + \frac{\partial f}{\partial x} dX_t = \frac{\partial f}{\partial t} dt + \frac{\partial f}{\partial x} \mu_t dt
\end{split}
\end{equation}

But if $\sigma_t \neq 0$, there is a diffusion term $\sigma_t dW_t$ inside $d X_t$. We have to keep terms of order $dt$ and $d W_t = \sqrt{d_t}$. And do not forget:
\begin{equation}
    (d W_t)^2 = dt
\end{equation}

The answer is called Ito's formula, the main result in the theory of stochastic calculus.

\begin{theorem}[Ito's formula]
    Assume that the process X has a stochastic differential given by
    \begin{equation}
        d X_t = \mu_t d_t + \sigma_t dW_t
    \end{equation}
    where $\mu$ and $\sigma$ are adapted processes, and let $f$ be a $C^{1, 2}$-function. Then $f(t, X_t)$ has a stochastic differential give by
    \begin{equation}
    \label{eq:IL}
        d f(t, X_t) = \left\{ \frac{\partial f}{\partial t}(t, X_t) + \mu_t \frac{\partial f}{\partial x}(t, X_t) + \sigma_t^2 \frac{\partial^2 f}{\partial x^2}(t, X_t) \right\} dt + \sigma_t \frac{\partial f}{\partial x}(t, X_t) d W_t
    \end{equation}
\end{theorem}

\begin{proof}
We do not give a full formal proof here, we only give a heuristic one. If we make a Taylor expansion including second-order terms we obtain\begin{equation}
\label{eq:TE}
    df = \frac{\partial f}{\partial t} dt + \frac{\partial f}{\partial x} d X_t + \frac{\partial^2 f}{\partial x^2} (d X_t)^2 + \frac{1}{2} \frac{\partial^2 f}{\partial t^2} (dt)^2 + \frac{\partial^2 f}{\partial t \partial x} dt d X_t
\end{equation}

By the definition of $d X_t$, we obtain
\begin{equation}
    (d X_t)^2 = \mu_t^2 (dt)^2 + 2 \mu_t \sigma_t (dt) (dW_t) + \sigma_t^2 (dW_t)^2
\end{equation}

Substitute into Eq.~\ref{eq:TE}, and notice that $(dt)^2$-term and $(dt)(dW_t)$-term are negligible compared to the $dt$-term $(dW_t)^2 = d_t$, plus the fact that $(d W_t)^2 = dt$, we have Eq.~\ref{eq:IL} as a result.
\end{proof}

Here we show how to derive $d \ln F_i(t)$ if $F_i (t)$ has the form of Eq.~\ref{eq:1}:
\begin{equation}
    \begin{split}
        d \ln F_i (t) = & \frac{1}{F_i (t)} d F_i (t) - \frac{1}{2} \frac{1}{F_i (t)^2} (d F_i(t))^2 \\
        = & \sigma_i (t) d Z_i^i (t) - \frac{1}{2F_i (t)^2} \sigma_i (t)^2 F_i (t)^2 (d Z_i^i (t))^2 \\
        = & \sigma_i (t) d Z_i^i (t) - \frac{\sigma_i (t)^2}{2} dt
    \end{split}
\end{equation}

\section{Proof of Theorem~\ref{thm:FMDL}}
\label{App:p1.2}

\renewcommand{\theequation}{{D}\arabic{equation}}
\renewcommand{\thefigure}{{D}\arabic{figure}}

We first introduce a central result for change of measure, \textbf{The Radon-Nikodym Theorem}.
\begin{theorem}[The Radon-Nikodym Theorem]
    Consider the measure space $(X, \mathcal{F}, \mu)$, where we assume that $\mu$ is finite, i.e. $\mu(X) < \infty$. Let $\nu$ be a measure on $(X, \mathcal{F})$ such that $\nu << \mu$ on $\mathcal{F}$. Then there exists a non-negative function $f: X \rightarrow R$ such that:\begin{equation}
        \begin{split}
            & f\; \textit{is}\; \mathcal{F}-measurable\\
            & \int_X f(x) d\mu(x) < \infty \\
            & \nu(A) = \int_A f(x) d \mu(X),\; \forall A \in \mathcal{F}.
        \end{split}
    \end{equation}
    
    The function $f$ is called the \textbf{Radon-Nikodym derivative} of $\nu$ w.r.t. $\mu$. It is uniquely determined $\mu$-almost everywhere and we write\begin{equation}
        f(x) = \frac{d\nu(x)}{d\mu(x)}
    \end{equation}
    or alternatively\begin{equation}
        d\nu(x) = f(x) d\mu(x)
    \end{equation}
\end{theorem}

The Radon-Nikodym theorem states that, under certain conditions, any measure $\nu$ can be expressed in this way with respect to another measure $\mu$ on the same space. We do not give out a proof here.

Another important theorem is "\textbf{Abstract Bayes' Formula}", which shows how conditional expected values under one probability measure are related to conditional expectations under another probability measure.

\begin{theorem}[Bayes' Formula]
    Assume that $X$ is a random variable on $(\Omega, \mathcal{F}, P)$, and let $Q$ be another probability measure on $(\Omega, \mathcal{F})$ with Radon-Nikodym derivative\begin{equation*}
        L = \frac{dQ}{dP} \; on \; \mathcal{F}.
    \end{equation*}
    
    Assume that $X \in L^1(\Omega, \mathcal{F}, Q)$ and that $\mathcal{G}$ is a sigma-algebra with $\mathcal{G} \subseteq \mathcal{F}$. Then\begin{equation}
        \mathbb{E}^Q[X|\mathcal{G}] = \frac{\mathbb{E}^P[L\cdot X|\mathcal{G}]}{\mathbb{E}^P[L|\mathcal{G}]}, \; Q-a.s.
    \end{equation}
\end{theorem}

\begin{proof}
    We show first that \begin{equation}
    \label{eq:BF}
        \mathbb{E}^Q[X|\mathcal{G}] \cdot \mathbb{E}^P[L|\mathcal{G}] = \mathbb{E}^P[L\cdot X|\mathcal{G}], \; P-a.s.
    \end{equation}
    
    We show this by proving that for an arbitrary $G \in \mathcal{G}$ the $P$-integral of both sides coincide. The left side is
    \begin{equation}
        \begin{split}
            \int_G \mathbb{E}^Q[X|\mathcal{G}] \cdot \mathbb{E}^P[L|\mathcal{G}] dP & = \int_G \mathbb{E}^P[L \cdot \mathbb{E}^Q[X|\mathcal{G}]|\mathcal{G}] dP \\
            & = \int_G L \cdot \mathbb{E}^Q[X|\mathcal{G}] dP \\
            & = \int_G \mathbb{E}^Q[X|\mathcal{G}] dQ \\
            & = \int_G X dQ
        \end{split}
    \end{equation}
    
    The first equality follows from the fact that $\mathbb{E}^Q[X|\mathcal{G}]$ is $\mathcal{G}$-measurable, and for a $\mathcal{G}$-measurable Y, $\mathbb{E}^P[XY|\mathcal{G}] = Y \mathbb{E}^P[X|\mathcal{G}]$. The second equality and the fourth equality is nothing but the definition of the conditional expectation. The third equality follows from Radon-Nikodym derivative.
    
    Integrating the right-hand side we obtain\begin{equation}
        \begin{split}
            \int_G \mathbb{E}^P[L \cdot X|\mathcal{G}] dP & = \int_G L \cdot X dP\\
            & = \int_G X dQ
        \end{split}
    \end{equation}
    
    Thus Eq.~\ref{eq:BF} holds $P$-a.s. and since $Q << P$, also $Q$-a.s. Now we show that $\mathbb{E}^P[L|\mathcal{G}] \neq 0$ $Q$-a.s.\begin{equation}
        \begin{split}
            Q(\mathbb{E}^P[L|\mathcal{G}] = 0) & = \int_{\{\mathbb{E}^P[L|\mathcal{G}] = 0\}} dQ\\
            & = \int_{\{\mathbb{E}^P[L|\mathcal{G}] = 0\}} L dP\\
            & = \int_{\{\mathbb{E}^P[L|\mathcal{G}] = 0\}} \mathbb{E}^P[L|\mathcal{G}] dP\\
            & = 0
        \end{split}
    \end{equation}
\end{proof}

\begin{theorem}
\label{thm:CEUEMM}
    Let $Q \in \mathcal{Q}$ be an EMM with numeraire $B$ and let $U$ be a $Q$-price process. Consider the probability measure $Q^U$ on $(\Omega. \mathcal{F})$ defined by
    \begin{equation}
        \frac{dQ^U}{dQ} = \frac{D(0, T)}{D^U(0, T)} = \frac{U_T B_0}{B_T U_0}
    \end{equation}
    where $D^U(t, T) = \frac{U_t}{U_T}$ is the $U$-discount factor. Then, for any $X \in L^1(\Omega, Q)$, we have
    \begin{equation}
        \mathbb{E}^Q[D(t, T) X | \mathcal{F}_t^W] = \mathbb{E}^{Q^U}[D^U(t, T) X | \mathcal{F}_t^W], \; t \in [0, T]
    \end{equation}
\end{theorem}

\begin{proof}
    We denote
    \begin{equation}
        Z_t \coloneqq \frac{D(0, T)}{D^U(0, T)} = \frac{U_T B_0}{B_T U_0}, \; t \in [0, T].
    \end{equation}
    
    Since $U$ is a $Q$-price process,
    \begin{equation}
    \begin{split}
        Z_t & = \frac{B_0}{U_0} \mathbb{E}^Q \left[ \frac{U_T}{B_T} | \mathcal{F}_t^W \right]\\
        & = \mathbb{E}^Q \left[ \frac{B_0}{U_0} \frac{U_T}{B_T} | \mathcal{F}_t^W \right] \\
        & = \mathbb{E}^Q \left[ \frac{D(0, T)}{D^U(0, T)} | \mathcal{F}_t^W \right]\\
        & = \mathbb{E}^Q \left[ Z_T | \mathcal{F}_t^W \right]
    \end{split}
    \end{equation}
    i.e. $Z$ is a strict $Q$-martingale. Then, by Bayes' formula,
    \begin{equation}
    \begin{split}
        \mathbb{E}^{Q^U}[D^U(t, T) X | \mathcal{F}_t^W] & = \frac{\mathbb{E}^{Q}[D^U(t, T) X Z_T | \mathcal{F}_t^W]}{\mathbb{E}^Q \left[ Z_T | \mathcal{F}_t^W \right]}\\
        & = \mathbb{E}^Q \left[ D^U(t, T) X \frac{Z_T}{Z_t} | \mathcal{F}_t^W \right]\\
        & = \mathbb{E}^Q \left[ D^U(t, T) X \frac{U_T B_0 B_t U_0}{B_T U_0 U_t B_0} | \mathcal{F}_t^W \right]\\
        & = \mathbb{E}^Q \left[ D^U(t, T) X \frac{D(t, T)}{D^U(t, T)} | \mathcal{F}_t^W \right]\\
        & = \mathbb{E}^Q \left[ D(t, T) X | \mathcal{F}_t^W \right]
    \end{split}
    \end{equation}
\end{proof}

From the proof we can see that $Q^U$ is an EMM with numeraire $U$. Corollary below shows how to compute the Radon-Nikodym derivative of two EMMs.

\begin{corollary}
\label{cor:RNDE}
    Let $U$, $V$ be $Q$-price processes with corresponding EMMs $Q^U$, $Q^V$, respectively. Then, we have
    \begin{equation}
        \frac{dQ^V}{dQ^U} | \mathcal{F}^W_t = \frac{V_t U_0}{U_t V_0}
    \end{equation}
\end{corollary}

\begin{proof}
    \begin{equation}
        \begin{split}
            \frac{dQ^V}{dQ^U} | \mathcal{F}^W_t & = \mathbb{E}^{Q^U} \left[ \frac{dQ^V}{dQ^U} | \mathcal{F}^W_t \right]\\
            & = \mathbb{E}^{Q^U} \left[ \frac{dQ^V}{dQ} \frac{dQ}{dQ^U} | \mathcal{F}^W_t \right]\\
            & = \mathbb{E}^{Q^U} \left[ \frac{V_T U_0}{U_T V_0} | \mathcal{F}^W_t \right]\\
            & = \frac{U_0}{V_0} \mathbb{E}^{Q^U} \left[ \frac{V_T}{U_T} | \mathcal{F}^W_t \right]\\
            & = \frac{U_0}{V_0} \mathbb{E}^{Q^V} \left[ \frac{V_t}{U_t} | \mathcal{F}^W_t \right]\\
            & = \frac{V_t U_0}{U_t V_0}
        \end{split}
    \end{equation}
    where the fifth equation uses Theorem~\ref{thm:CEUEMM} twice.
\end{proof}

If we consider a fixed time interval $[0, T]$, a probability space $(\Omega, \mathcal{F}, P)$ with some filtration $\{\mathcal{F}_t\}_{t \geq 0}$, an adapted $d$-dimensional Brownian motion $W = (W^1, ...,W^d)$, a vector process $h = (h^1, ...,h^d)$ which is "integrable enough", a real number $x_0$, and define the process $M$ by
\begin{equation}
\label{eq:M}
    M_t = x_0 + \sum\limits_{i=1}^d \int_0^t h^i_s dW^i_s, \; t \in [0,T],
\end{equation}
then $M$ is a martingale. Now we want to ask if every $\mathcal{F}_t$-adapted martingale $M$ can be written in the form of Eq.~\ref{eq:M}. The following theorem shows, under some reasonable conditions, the answer is yes.

\begin{theorem}[The Martingale Representation Theorem]
    Let $W$ be a $d$-dimensional Wiener process, and assume that the flitration $\mathcal{F}$ is defined as
    \begin{equation}
        \mathcal{F}_t = \mathcal{F}_t^W, \; t\in [0,T].
    \end{equation}
    
    Let $M$ be any $\mathcal{F}_t$-adapted martingale. Then there exist uniquely determined $\mathcal{F}_t$-adapted processes $h^1, ...,h^d$ such that $M$ has the representation
    \begin{equation}
        M_t = M_0 + \sum\limits_{i=1}^d \int_0^t h^i_s dW^i_s, \; t \in [0,T].
    \end{equation}
    
    If the martingale $M$ is square integrable, then $h^1, ...,h^d$ are in $\mathbb{L}^2$
\end{theorem}

We do not give out a proof here.

We want to find out the effect of change of measure will have upon a Brownian motion. For any Radon-Nikodym derivative $L_t = \frac{dQ}{dP}$, from the earlier discussion we know that $L_t$ is a non-negative $P$-martingale. It is then natural to define $L$ as
\begin{equation}
\label{eq:L}
\begin{split}
    d L_t & = -\lambda_t L_t dW_t\\
    L_0 & = 1\\
    \mathbb{E}^P [L_t] & = L_0 = 1,\; t \in[0, T]\\
\end{split}
\end{equation}

Again using Ito's Lemma, we know that we can express $L$ as
\begin{equation}
    L_t =  e^{-\int_0^T \lambda_s d W_s - \frac{1}{2} \int_0^T \lambda_s^2 ds}
\end{equation}

The following important theorem tells us how to write a $P$-Wiener process $W$ under new probability measure $Q$.

\begin{theorem}[Girsanov's Theorem]
    Let $W$ be a $d$-dimensional standard $P$-Wiener process on $(\Omega, \mathcal{F}, P)$ and let $\lambda$ be any $d$-dimensional adapted column vector process. Define the process $L$ as in Eq.~\ref{eq:L}, and define the new probability measure $Q$ on $\mathcal{F}_T$ by
    \begin{equation}
        L_T = \frac{dQ}{dP}
    \end{equation}
    Then the process $W^\lambda$ defined by
    \begin{equation}
        W^\lambda_t = W_t + \int_0^t \lambda_s ds \;, \; t \in [0, T],
    \end{equation}
    is a Wiener process on $(\Omega, \mathcal{F}, Q, \mathcal{F}_t)$.
\end{theorem}

We do not give out a proof here.

The following theorem shows how correlated Wiener process behave under change of measure.

\begin{theorem}
\label{thm:CDC}
    Let $W$ be a $d$-dimensional correlated $P$-Wiener process on $(\Omega, \mathcal{F}, P)$ and let $\lambda$ be any $d$-dimensional adapted column vector process.Define the process $L$ as in Eq.~\ref{eq:L}, and define the new probability measure $Q$ on $\mathcal{F}_T$ by
    \begin{equation}
        L_T = \frac{dQ}{dP}
    \end{equation}
    Then the process $W^\lambda$ defined by
    \begin{equation}
    \label{eq:WWC}
        dW^\lambda_t = dW_t + \rho \lambda_t dt \;,
    \end{equation}
    is a Wiener process on $(\Omega, \mathcal{F}, Q, \mathcal{F}_t)$ with correlation matrix $\rho$.
\end{theorem}

\begin{proof}
    By the martingale representation theorem, there exists a unique ($P$-a.s.) $d$-dimensional process $\bar{\lambda}$ such that
    \begin{equation}
    \label{eq:DLT}
        d L_t = - L_t \bar{\lambda}_t \cdot d\bar{W}_t
    \end{equation}
    under standard $P$-Wiener process $\bar{W}$. Eq.~\ref{eq:DLT} can be further written as
    \begin{equation}
    \begin{split}
        d L_t & = - L_t \bar{\lambda}_t \cdot d\bar{W}_t\\
        & = -L_t \bar{\lambda}_t \cdot (A^{-1} dW_t)\\
        & = -L_t ((A^{-1})^* \bar{\lambda}_t) \cdot d W_t\\
        & = -L_t \lambda_t \cdot dW_t
    \end{split}
    \end{equation}
    where $\lambda_t \equiv (A^{-1})^* \bar{\lambda}_t$. By Girsanov's theorem, we have a standard $Q$-Wiener process defined by
    \begin{equation}
    \label{eq:BW}
        d\bar{W}^{\bar{\lambda}}_t = d\bar{W}_t + \bar{\lambda}_t dt.
    \end{equation}
    
    Multiply both sides of Eq.~\ref{eq:BW} by $A$ and define $dW_t^{\lambda} \coloneqq A d\bar{W}^{\bar{\lambda}}_t$, we have Eq.~\ref{eq:WWC}.
\end{proof}

Now we prove Theorem~\ref{thm:FMDL}.
\begin{proof}
    We want to find the deterministic functions $\mu_k^i(t, F(t))$, where $F(t) = (F_1(t), ..., F_M(t))$, that satisfies
    \begin{equation}
    \label{eq:OF}
        d F_k(t) = \mu_k^i(t, F(t)) F_k(t) dt + \sigma_k (t) F_k(t) dZ_k^i (t), \; k \neq i.
    \end{equation}
    
    From Corollary~\ref{cor:RNDE}, the Radon-Nikodym derivative of $Q^{i-1}$ w.r.t. $Q^i$ at time $t$ is
    \begin{equation}
        \frac{dQ^{i-1}}{dQ^i} | \mathcal{F}^W_t = \frac{p(t, T_{i - 1}) p(0, T_i)}{p(t, T_i) p(0, T_{i-1})}
    \end{equation}
    which we denote as $\gamma_t^i$. Using Eq.~\ref{eq:FR} we can write
    \begin{equation}
        \gamma_t^i = \frac{p(0, T_i)}{p(0, T_{i-1})}(F_i(t) \tau_i + 1).
    \end{equation}
    
    The $Q^i$-dynamics of $\gamma^i$ is given by
    \begin{equation}
    \label{eq:BP}
    \begin{split}
        d\gamma^i_t & = \frac{p(0, T_i)}{p(0, T_{i-1})} d F_i(t) \tau_i\\
        & = \frac{p(0, T_i)}{p(0, T_{i-1})} \tau_i \sigma_i(t) F_i(t) dZ^i_i(t)\\
        & = \frac{\gamma_t^i}{F_i(t) \tau_i + 1} \tau_i \sigma_i(t) F_i(t) dZ^i_i(t)
    \end{split}
    \end{equation}
    last equality uses Eq.~\ref{eq:1}. Using Theorem~\ref{thm:CDC}, we can have
    \begin{equation}
        d Z^i (t) = dZ^{i - 1}(t) -\rho \lambda dt,
    \end{equation}
    with component
    \begin{equation}
        d Z^i_h(t) = dZ^{i - 1}_h(t) + \rho^{hi} \frac{\tau_i \sigma_i (t) F_i(t)}{1 + F_i(t) \tau_i} dt
    \end{equation}
    
    Applying this inductively we obtain
    \begin{equation}
    \label{eq:ZI}
        \begin{split}
            d Z^i_h(t) = dZ^{k}_h(t) + \sum\limits_{j=k+1}^i \frac{\tau_j \rho_{k, j} \sigma_k(t) \sigma_j(t) F_j(t)}{1+\tau_j F_j(t)} dt, \;&\; k < i\\
            d Z^i_h(t) = dZ^{k}_h(t) - \sum\limits_{j=i+1}^k \frac{\tau_j \rho_{k, j} \sigma_k(t) \sigma_j(t) F_j(t)}{1+\tau_j F_j(t)} dt, \;&\; k > i.
        \end{split}
    \end{equation}
    Substitute Eq.~\ref{eq:ZI} into Eq.~\ref{eq:OF} and equating the $Q^k$-drift to zero, we find $\mu_k^i(t, F(t))$:
    \begin{equation}
        \begin{split}
            \mu_k^i(t, F(t)) & = -\sum\limits_{j=k+1}^i \frac{\tau_j \rho_{k, j} \sigma_k(t) \sigma_j(t) F_j(t)}{1+\tau_j F_j(t)},\; \; k < i\\
            \mu_k^i(t, F(t)) & = \sum\limits_{j=i+1}^k \frac{\tau_j \rho_{k, j} \sigma_k(t) \sigma_j(t) F_j(t)}{1+\tau_j F_j(t)},\; \; k > i\\
        \end{split}
    \end{equation}
    
\end{proof}

\section{Proof of Theorem~\ref{thm:ER}}
\label{App:pER}
\renewcommand{\theequation}{{E}\arabic{equation}}
\renewcommand{\thefigure}{{E}\arabic{figure}}

\begin{proof}
    For $i = M$ it is trivial to show that
    \begin{equation}
    \label{eq:TK}
        d F_M = \sigma_M(t) F_M(t) dZ^M_M(t).
    \end{equation}
    
    Eq.~\ref{eq:TK} is just GBM with $\sigma_N$ bounded. Thus a solution does exist. Now we prove the rest part by induction. Assume now that Eq.~\ref{eq:PF} admits a solution for $i+1, ..., M$, then we can write the $i^{th}$ component of Eq.~\ref{eq:PF} as
    \begin{equation}
        d F_i(t) = \mu_i (t, F_{i+1}(t), ..., F_M(t))F_i(t) dt + \sigma_i(t) F_i(t) d Z_i^M(t),
    \end{equation}
    where the point is that $\mu_i$ only depend on $F_k$ for $k = i+1, ..., M$ and not on $F_i$. Denoting the vector $(F_{i+1}, ..., F_M)'$ by $F_{i+1}^M$, we can solve the above SDE using Ito's Lemma:
    \begin{equation}
        F_i(t) = F_i(0) e^{\int_0^t \mu_i(s, F_{i+1}^M(s)) - \frac{\sigma_i(s)^2}{2} ds + \int_0^t \sigma_i(s) d Z^i_i(s)},
    \end{equation}
    for $0 \leq t \leq T_{i-1}$. This proves existence. It also follows by induction that all LIBOR rate processes will be positive given an initial positive LIBOR term structure. Thus the process as shown in Eq.~\ref{eq:BP} is bounded and consequently satisfies the Novikov condition.
\end{proof}

\section{Discretization of Probability Distributions}
\renewcommand{\theequation}{{F}\arabic{equation}}
\renewcommand{\thefigure}{{F}\arabic{figure}}
\label{App:d}

\begin{figure}[ht]
\centering
\includegraphics[width=0.6\textwidth]{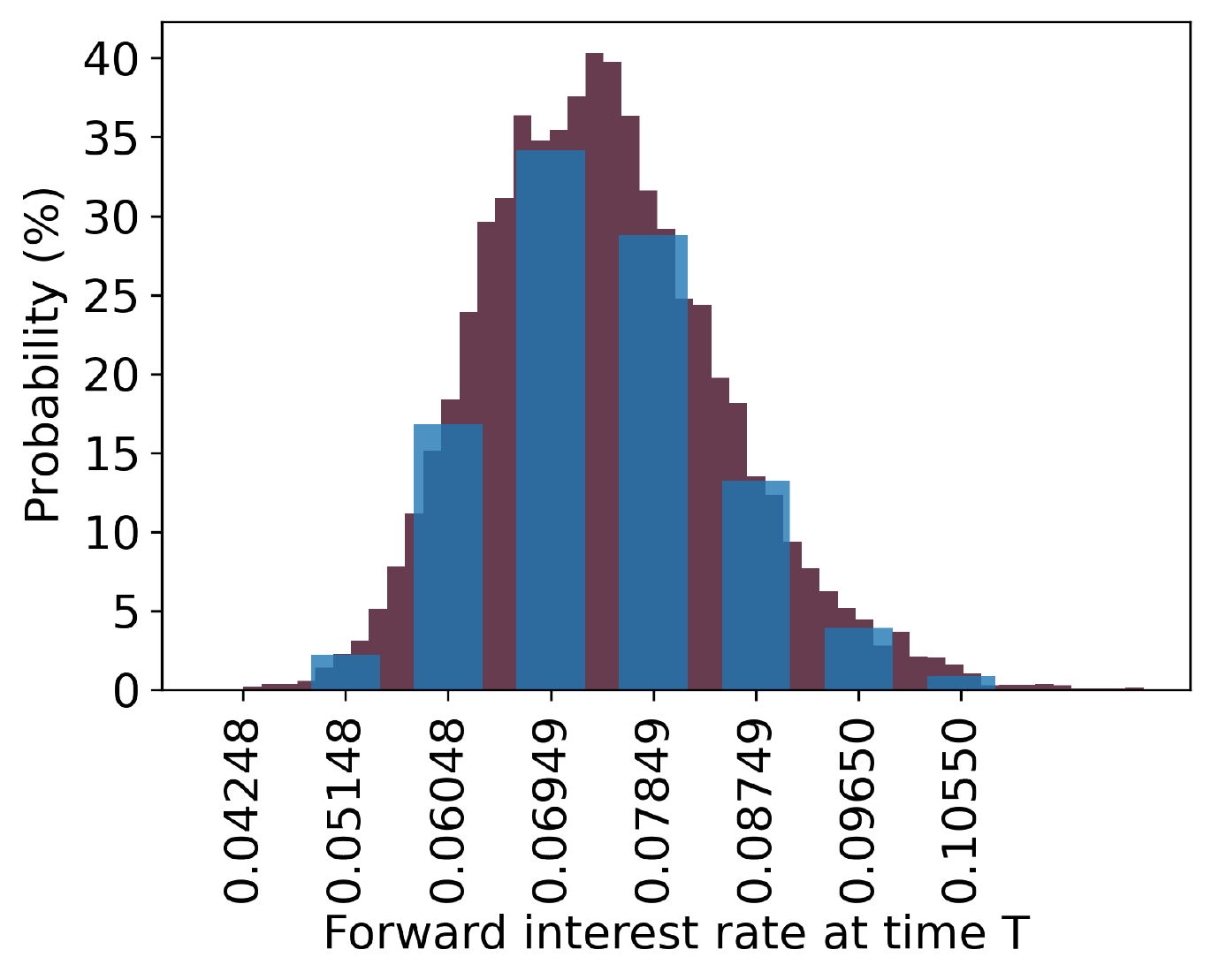}
\caption{Comparison of Discretized Lognormal Distribution and True Lognormal Distribution}
\label{fig:dc}
\end{figure}

Fig.~\ref{fig:CI} (b) depicted the discretized lognormal distribution (in blue) with $n = 3$ qubits and the true lognormal distribution (in eggplant color, with 10000 simulations divided into 50 bins) at time $T = 10$. The figure only shows one possible probability distribution when $T = 10$ because the interest rate value at $T = 9$ is generated using Monte Carlo simulation. As illustrated in the figure above, true lognormal distribution exhibits more outliers on the right arm, whereas discretized lognormal distribution favor the left arm slightly more. The mean and skewness statistics for discretized and true distribution are $7.391 \times 10^{-2}$, $9.330 \times 10^{-4}$ and $7.406 \times 10^{-2}$, $0.4528$ respectively. The mean of the discretized lognormal probability distribution is slightly less than that of the true distribution, but the skewness is significantly less than that of true distribution. This can lead to the outcome illustrated in Fig.\ref{fig:CI} (b), where the Quantum-classic Hybrid Method Outcome is consistently lower than the theoretical value. This is exacerbated when initial interest rates are high, as they are when $T$ is large.

\end{appendix}


\begin{thebibliography}{10}
\renewcommand{\bibnumfmt}[1]{#1.}


\bibitem{Hull2003}
\bibinfo{author} {Hull, J. C.} 
\newblock \emph{\bibinfo{Book}{Options futures and other derivatives.}}
\bibinfo{Publisher}{Pearson Education India}
  (\bibinfo{year}{2003}).

\bibitem{Tuckman2012}
\bibinfo{author} {Tuckman, B., \& Serrat, A.} 
\newblock \emph{\bibinfo{Book}{Fixed income securities: tools for today's markets, 3rd Edition.}}
\bibinfo{Publisher}{John Wiley \& Sons}
  (\bibinfo{year}{2012}).

\bibitem{Chacko2016}
\bibinfo{author} {Chacko, G., Sj\"oman, A., Motohashi, H., \& Dessain, V} 
\newblock \emph{\bibinfo{Book}{Credit Derivatives, Revised Edition: A Primer on Credit Risk, Modeling, and Instruments.}}
\bibinfo{Publisher}{Pearson Education}
  (\bibinfo{year}{2016}).

\bibitem{Black1973}
\bibinfo{author} {Black, F. \& Scholes, M.} 
\bibinfo{title} {The pricing of options and corporate liabilities.}
\newblock \emph{\bibinfo{journal}{J. Political Econ.}}
\textbf{\bibinfo{volume}{81}}, \bibinfo{pages}{637–654}
  (\bibinfo{year}{1973}).

\bibitem{Merton1973}
\bibinfo{author} {Merton, R.C.} 
\bibinfo{title} {Theory of rational option pricing. }
\newblock \emph{\bibinfo{journal}{The Bell Journal of economics and management science}}
\textbf{\bibinfo{volume}{4}}, \bibinfo{pages}{141-183}
  (\bibinfo{year}{1973}).


\bibitem{EUreport2021}
\bibinfo{title} {ESMA Annual Statistical Report on EU Derivatives Markets 2021. }
\bibinfo{pages}{Retrieved from: https://www.esma.europa.eu/}
 (\bibinfo{year}{2021}).
  

\bibitem{Heath1990}
\bibinfo{author} {Heath, D., Jarrow, R., \& Morton, A.} 
\bibinfo{title} {Bond Pricing and the Term Structure of Interest Rates: A Discrete Time Approximation.}
\newblock \emph{\bibinfo{journal}{Journal of Financial and Quantitative Analysis}}
\textbf{\bibinfo{volume}{25}}, \bibinfo{pages}{419-440}
  (\bibinfo{year}{1990}).

\bibitem{Heath1991}
\bibinfo{author} {Heath, D., Jarrow, R., \& Morton, A.} 
\bibinfo{title} {Contingent Claims Valuation with a Random Evolution of Interest Rates.}
\newblock \emph{\bibinfo{journal}{Review of Futures Markets}}
\textbf{\bibinfo{volume}{9}}, \bibinfo{pages}{54-76}
  (\bibinfo{year}{1991}).

\bibitem{Heath1992}
\bibinfo{author} {Heath, D., Jarrow, R., \& Morton, A.} 
\bibinfo{title} {Bond pricing and the term structure of interest rates: A new methodology for contingent claims valuation.}
\newblock \emph{\bibinfo{journal}{Econometrica: Journal of the Econometric Society}}
\textbf{\bibinfo{volume}{60}}, \bibinfo{pages}{77-105}
  (\bibinfo{year}{1992}).


\bibitem{Brace1997}
\bibinfo{author} {Brace, A., G\c atarek, D., \& Musiela, M. } 
\bibinfo{title} {The market model of interest rate dynamics.}
\newblock \emph{\bibinfo{journal}{Mathematical Finance}}
\textbf{\bibinfo{volume}{7}}, \bibinfo{pages}{127-155}
  (\bibinfo{year}{1997}).


\bibitem{Huang2014}
\bibinfo{author} {Huang, J. S.} 
\emph{\bibinfo{title} {A Libor Market Model Approach for Measuring Counterparty Credit Risk Exposure.}}
 \bibinfo{pages}{Master thesis, University of Amsterdam}
 (\bibinfo{year}{2014}).

\bibitem{Kajsajuntti2004}
\bibinfo{author} {Kajsajuntti, L.} 
\emph{\bibinfo{title} {Pricing of Interest Rate Derivatives with the LIBOR Market Model.}}
 \bibinfo{pages}{Master thesis, Royal Institute of Technology}
 (\bibinfo{year}{2004}).

\bibitem{Riga2011}
\bibinfo{author} {Riga, C.} 
\emph{\bibinfo{title} {The Libor Market Model: from theory to calibration.}}
 \bibinfo{pages}{Master thesis, The University of Bologna}
 (\bibinfo{year}{2011}).

\bibitem{Xiong2013}
\bibinfo{author} {Xiong, C. W.} 
\emph{\bibinfo{title} {Introduction to Interest Rate Models.}}
 \bibinfo{pages}{Notes, retrieved from: https:$//$modelmania.github.io}
 (\bibinfo{year}{2013}).

\bibitem{Pena2010}
\bibinfo{author} {Pena, A., di Sabatino, A., Ligato, S., Ventura, S.,\& Bertagna, A.} 
\emph{\bibinfo{title} {The One Factor Libor Market Model Using Monte Carlo Simulation: An Empirical Investigation.}}
 \bibinfo{pages}{Working paper. SDA Bocconi School of Management, Thomson Reuters, Banca Carige, and Unicredit Group.}
 (\bibinfo{year}{2010}).
   
\bibitem{Brigo2006}
\bibinfo{author} {Brigo D., \& Mercurio, F.} 
\newblock \emph{\bibinfo{journal}{Interest Rate Models: Theory and Practice - with Smile, Inflation and Credits.}}
\bibinfo{pages}{Heidelberg, Springer}
  (\bibinfo{year}{2006})





\bibitem{Orus2019}
\bibinfo{author} {Or\'us, R., Mugel, S., \& Lizaso, E.} 
\bibinfo{title} {Quantum computing for finance: Overview and prospects.}
\newblock \emph{\bibinfo{journal}{Rev. in Phys.}}
\textbf{\bibinfo{volume}{4}}, \bibinfo{pages}{100028}
  (\bibinfo{year}{2019}).



\bibitem{Baaquie2007}
\bibinfo{author} {Baaquie, B. E.} 
\newblock \emph{\bibinfo{Book}{Quantum finance: Path integrals and Hamiltonians for options and interest rates. }}
\bibinfo{Publisher}{Cambridge University Press}
  (\bibinfo{year}{2007}).

\bibitem{Zhang2010}
\bibinfo{author} {Zhang, C., \& Huang, L. } 
\bibinfo{title} {A quantum model for the stock market. }
\newblock \emph{\bibinfo{journal}{Physica A: Statistical Mechanics and its Applications}}
\textbf{\bibinfo{volume}{389}}, \bibinfo{pages}{5769-5775}
  (\bibinfo{year}{2010}).

\bibitem{Meng2016}
\bibinfo{author} {Meng, X., Zhang, J. W., \& Guo, H.} 
\bibinfo{title} {Quantum Brownian motion model for the stock market.}
\newblock \emph{\bibinfo{journal}{Physica A: Statistical Mechanics and its Applications}}
\textbf{\bibinfo{volume}{452}}, \bibinfo{pages}{281-288}
  (\bibinfo{year}{2016}).

\bibitem{Mugel2022}
\bibinfo{author} {Mugel, S., Carlos Kuchkovsky, C., Sanchez, E.,  Fernandez-Lorenzo, S., Luis-Hita, J., Lizaso, E., \& Orus, R.} 
\bibinfo{title} {Dynamic portfolio optimization with real datasets using quantum processors and quantum-inspired tensor networks.}
\newblock \emph{\bibinfo{journal}{Physical Review Research}}
\textbf{\bibinfo{volume}{4}}, \bibinfo{pages}{013006}
  (\bibinfo{year}{2022}).
  
\bibitem{Hegade2021}
\bibinfo{author} {Hegade, N. H., Chandarana, P., Paul, K., Chen, X., Albarrán-Arriagada, F., \& Solano, E.} 
\bibinfo{title} {Portfolio optimization with digitized-counterdiabatic quantum algorithms.}
\newblock \emph{\bibinfo{journal}{arXiv preprint}}
\bibinfo{pages}{arXiv:2112.08347}
  (\bibinfo{year}{2021}).

  
 \bibitem{Stamatopoulos2021}
\bibinfo{author} {Stamatopoulos, N., Mazzola, G., Woerner, S. \& Zeng, W. J.} 
\bibinfo{title} {Towards Quantum Advantage in Financial Market Ris.k using Quantum Gradient
Algorithms.}
\newblock \emph{\bibinfo{journal}{arXiv preprint}}
 \bibinfo{pages}{arXiv:2111.12509}
  (\bibinfo{year}{2021}). 

 \bibitem{Coyle2021}
\bibinfo{author} {Coyle, B., Henderson, M., Le, J.C.J., Kumar, N., Paini, M. \& Kashefi, E.} 
\bibinfo{title} {Quantum versus classical generative modelling in finance.}
\newblock \emph{\bibinfo{journal}{Quantum Science and Technology}}
\textbf{\bibinfo{volume}{6}}, \bibinfo{pages}{024013}
  (\bibinfo{year}{2021}).
  
\bibitem{Rebentrost2018}
\bibinfo{author} {Rebentrost, P., Gupt,B., \& Bromley, T. B.} 
\bibinfo{title} {Quantum computational finance: Monte Carlo pricing of financial derivatives.}
\newblock \emph{\bibinfo{journal}{Phys. Rev. A}}
\textbf{\bibinfo{volume}{98}}, \bibinfo{pages}{022321}
  (\bibinfo{year}{2018})

\bibitem{Martin2019}
\bibinfo{author} {Martin, A., Candelas, B., Rod\'iguez-Rozas, A., Mart\'in-Guerrero, J. D., Chen,X., Lamata, L., Or\'us, R., Solano,E., \& Sanz, M., } 
\bibinfo{title} {Towards Pricing Financial Derivatives with an IBM Quantum Computer.}
\newblock \emph{\bibinfo{journal}{Physical Review Research}}
\textbf{\bibinfo{volume}{3}}, \bibinfo{pages}{013167}
  (\bibinfo{year}{2021}).

\bibitem{Woerner2019}
\bibinfo{author} {Woerner, S., \& Egger, D. J.} 
\bibinfo{title} {Quantum risk analysis.}
\newblock \emph{\bibinfo{journal}{npj Quantum Info.}}
\textbf{\bibinfo{volume}{5}}, \bibinfo{pages}{15}
  (\bibinfo{year}{2019}).

\bibitem{Zoufal2019}
\bibinfo{author} {Zoufal, C., Lucchi, A., \& Woerner, S.} 
\bibinfo{title} {Quantum Generative Adversarial Network for Learning and Loading Random Distributions.}
\newblock \emph{\bibinfo{journal}{npj Quantum Info.}}
\textbf{\bibinfo{volume}{5}}, \bibinfo{pages}{103}
  (\bibinfo{year}{2019}).

\bibitem{Stamatopoulos2019}
\bibinfo{author} {Stamatopoulos, N., Egger, D. J., Sun, Y., Zoufal, C., Iten, R., Shen, N., \& Woerner, S.} 
\bibinfo{title} {Option Pricing using Quantum Computers.}
\newblock \emph{\bibinfo{journal}{Quantum}}
\textbf{\bibinfo{volume}{4}}, \bibinfo{pages}{291}
  (\bibinfo{year}{2020}).

\bibitem{Egger2019}
\bibinfo{author} {Egger, D. J., Guti\'errez, R. G., Mestre, J. C., \& Woerner, S.} 
\bibinfo{title} {Credit Risk Analysis using Quantum Computers.}
\newblock \emph{\bibinfo{journal}{IEEE Transactions on Computers}}
\textbf{\bibinfo{volume}{70}}, \bibinfo{page}{2136-2145}
  (\bibinfo{year}{2020}).

\bibitem{Tang2020b}
\bibinfo{author} {Tang, H., Pal, A., Qiao, L. F., Wang, T. Y., Gao, J. \& Jin, X. M.} 
\bibinfo{title} {Quantum Computation for Pricing the Collateralized Debt Obligations.}
\newblock \emph{\bibinfo{journal}{Quantum Engineering}}
\bibinfo{page}{3,e84}
  (\bibinfo{year}{2021}).
  
\bibitem{Miyamoto2022}
\bibinfo{author} {Miyamoto, K.} 
\bibinfo{title} {Bermudan option pricing by quantum amplitude estimation and Chebyshev interpolation.}
\newblock \emph{\bibinfo{journal}{EPJ Quantum Technology}}
\textbf{\bibinfo{volume}{9}}, \bibinfo{pages}{3}
  (\bibinfo{year}{2022})
  
\bibitem{Miyamoto2022b}
\bibinfo{author} {Miyamoto, K., \& Kubo, K.} 
\bibinfo{title} {Pricing Multi-Asset Derivatives by Finite-Difference Method on a Quantum Computer.}
\newblock \emph{\bibinfo{journal}{IEEE Transactions on Quantum Engineering}}
\textbf{\bibinfo{volume}{3}}, \bibinfo{pages}{1-25}
  (\bibinfo{year}{2022})

\bibitem{Brassard2002}
\bibinfo{author} {Brassard, G., Hoyer,  P., Mosca, M., \& A. Tapp, A.} 
\bibinfo{title} {Quantum Amplitude Amplification and Estimation.}
\newblock \emph{\bibinfo{journal}{Contemporary Mathematics}}
\textbf{\bibinfo{volume}{305}}, \bibinfo{pages}{53-74}
  (\bibinfo{year}{2002})

\bibitem{Qiskit2019}
\bibinfo{author} {Aleksandrowicz, G., et al.} 
\bibinfo{title} {Qiskit: An open-source framework for quantum computing.}
\bibinfo{pages}{10.5281/zenodo.2562110}
  (\bibinfo{year}{2019}).


\bibitem{Girsanov1960}
\bibinfo{author} {Girsanov, I. V.} 
\bibinfo{title} {On transforming a certain class of stochastic processes by absolutely continuous substitution of measures.}
\newblock \emph{\bibinfo{journal}{Theory of Probability and Its Applications}}
\textbf{\bibinfo{volume}{5}}, \bibinfo{pages}{285–301}
  (\bibinfo{year}{1960})


  \bibitem{Chakrabarti2021}
\bibinfo{author} {Chakrabarti, S., Krishnakumar, R., Mazzola, G., Stamatopoulos, N., Woerner, S., \& Zeng, W. J.} 
\bibinfo{title} {A threshold for quantum advantage in derivative pricing.}
\newblock \emph{\bibinfo{journal}{Quantum}}
\textbf{\bibinfo{volume}{5}}, \bibinfo{issue}{463}
  (\bibinfo{year}{2021}).
  
  
\bibitem{Jabbour2011}
\bibinfo{author} {Jabbour, G., \& Liu, Y.} 
\bibinfo{title} {Option Pricing And Monte Carlo Simulations.}
\newblock \emph{\bibinfo{journal}{Journal of Business \& Economics Research (JBER)}}
\textbf{\bibinfo{volume}{3}}, \bibinfo{issue}{9}
  (\bibinfo{year}{2011}).

  

\bibitem{Zhu2021}
\bibinfo{author} {Zhu, E. Y., Johri, S., Bacon, D., Esencan, M., Kim, J., et al.} 
\bibinfo{title} {Generative quantum learning of joint probability distribution functions.}
\newblock \emph{\bibinfo{journal}{arXiv preprint}}
\bibinfo{issue}{arxiv:2109.06315}
  (\bibinfo{year}{2021}).

\bibitem{Zhu2022}
\bibinfo{author} {Zhu, D. W., Shen, W. W., Gianib, A., Majumderb, S. R., Neculaesb, B., \& Johria, S.} 
\bibinfo{title} {Copula-based Risk Aggregation with Trapped Ion Quantum Computers.}
\newblock \emph{\bibinfo{journal}{arXiv preprint}}
\bibinfo{issue}{arxiv:2206.11937}
  (\bibinfo{year}{2022}).

\bibitem{Kiss2022}
\bibinfo{author} {Kiss, O., Grossi, M., Kajomovitz, E., \& Vallecorsa, S.} 
\bibinfo{title} {Conditional Born machine for Monte Carlo events generation.}
\newblock \emph{\bibinfo{journal}{arXiv preprint}}
\bibinfo{issue}{arxiv:2205.07674}
  (\bibinfo{year}{2022}).




\end{thebibliography}
\end{document}